\documentclass[11pt,reqno]{amsart}
\usepackage{epsfig}
\usepackage{graphicx}
\begin{document}
\baselineskip=0.5cm \numberwithin{equation}{section}
\theoremstyle{plain}
\newtheorem{thm}{Theorem}[section]
\newtheorem{lem}{Lemma}[section]
\newtheorem{prop}{Proposition}[section]
\newtheorem{coll}{Conclusion}
\theoremstyle{remark}
\newtheorem{rem}{Remark}[section]
\title {New variable separation approach: application to nonlinear diffusion equations}
\author{Shun-li Zhang$^{1,2}$ , Sen-yue Lou$^{1,3}$ and Chang-zheng Qu$^{2}$}
\dedicatory{{$^1$Physics Department of Shanghai Jiao Tong
University, Shanghai, 200030, P. R. China}\\ {$^{2}$Mathematics
Department of Northwest
University, Xi'an 710069, P. R. China}\\
{$^3$Physics Department of Ningbo University, Ningbo, 315211, P.
R. China}}
\begin{abstract}
The concept of the derivative-dependent functional separable
solution, as a generalization to the functional separable
solution, is proposed. As an application, it is used to discuss
the generalized nonlinear diffusion equations based on the
generalized conditional symmetry approach. As a consequence, a
complete list of canonical forms for such equations which admit
the derivative-dependent functional separable solutions is
obtained and some exact solutions to the resulting equations are
described.
\end{abstract}

\maketitle

\section{Introduction}
A number of methods have been proved to be effective for finding
symmetry reduction and constructing exact solutions to nonlinear
diffusion equations. These include the Lie's classical approach
\cite{Ames}, the nonclassical approach \cite{Blu}, the direct
method \cite{CK},\ the modified direct method \cite{TL}, the
generalized conditional symmetry (GCS) method \cite{Fok}, the
nonlocal symmetry method \cite{ISA}, the truncated Painlev\'e
approach \cite{New}, the sign-invariant and invariant space
methods \cite{Ga1}, the transformation method \cite{Ki1} and the
ansatz-based methods [10-13] etc. There are many different
directions of the mathematical and physical theory to concern
their exact solutions and various properties.

It is well known that the method of variable separation is one of
the most universal and efficient means for study of linear partial
differential equations (PDEs). Several methods of variable
separation for nonlinear partial differential equations (PDEs)
such as the classical method \cite{Miller}, the differential
St\"ackel matrix approach \cite{Ka1}, the ansatz-based method
[12-16], the geometrical method \cite{Do1}, the formal variable
separation approach (nonlinearization of the Lax pais or symmetry
constraints) \cite{Lou3} and the informal variable separation
methods \cite{Lou1} have been suggested. From the point of view of
symmetry group and the ansatz of the solution form, we now
emphasize two of those ansatzs. One is the ordinary additive or
product separable solution. The other is the functional separable
solution which is a generalization of the former, where the
compatibility of the symmetry constraint with the considered
equations is concerned [11-13]. In \cite{Lou1}, exact solutions
depending on arbitrary functions and their derivatives to many
$(2+1)$-dimensional nonlinear integrable models have emerged
through the variable separation process, and abundant localized
excitations and their rich interaction behaviors have been
revealed. All these prompt us to extend our results in
\cite{Zh1,Zh2} to be more general.

In \cite{Zh1,Zh2}, we have discussed the functional separable
solution \begin{eqnarray} f(u)=a(x)+b(t),
 \label{01}
\end{eqnarray}
to the generalized porous medium equation
\begin{eqnarray}
u_t=(D(u)u_x^n)_x+F(u),
 \label{02}
\end{eqnarray}
for $n=1$ and $n\not=1$ respectively by using the GCS method.

To obtain more abundant exact solutions to nonlinear PDEs, our
basic idea is to weaken the general symmetry constraint condition
and to include much more solutions. In this paper, we extend the
concept of functional separable solution (\ref{01}) to that of
derivative-dependent functional separable solution (DDFSS)
\begin{eqnarray}
f(u,u_x)=a(x)+b(t), \label{03}
\end{eqnarray}
and apply it to the nonlinear diffusion equation
\begin{eqnarray}
u_t=A(u,u_x)u_{xx}+B(u,u_x). \label{04}
\end{eqnarray}
It is clear that when $f_{u_x}(u,u_x)=0$, (\ref{03}) becomes
(\ref{01}), so we assume that $f_{u_x}(u,u_x)\not=0$ hereafter.

The compatibility of (\ref{03}) and (\ref{04}) can be described in
terms of the GCS method. The GCS is a natural generalization of
both the generalized symmetry and the conditional symmetry
\cite{Fok}.

Consider the general $m$-th order $(1+1)$-dimensional evolution
equation
\begin{eqnarray}
u_t=E(t,x,u,u_1,u_2,\cdots,u_m), \label{05}
\end{eqnarray}
where $u_k=\frac{\partial^k{u}}{\partial{x^k}},1\le{k}\le{m}$, and
$E$ is a smooth function of the indicated variables. Let
\begin{eqnarray}
V=\eta(t,x,u,u_1,u_2,\cdots,u_j)\frac{\partial}{\partial{u}},
\label{06}
\end{eqnarray}
be an evolutionary vector field and $\eta$ its characteristic.\\
{\bf Definition 1.} The evolutionary vector field (\ref{06}) is
said to be a generalized symmetry of (\ref{05}) if and only if
\begin{eqnarray}
V^{(m)}(u_t-E)|_L=0.\nonumber
\end{eqnarray}
where $L$ is the solution set of (\ref{05}), and $V^{(m)}$ is the
$m$-th
prolongation of $V$.\\
{\bf Definition 2.} The evolutionary vector field (\ref{06}) is
said to be a GCS of (\ref{05}) if and only if
\begin{eqnarray}
V^{(m)}(u_t-E)|_{{L}\bigcap{W}}=0. \label{07}
\end{eqnarray}
where $W$ is the set of equations $D^i_x{\eta}=0, i=0,1,2,\cdots.$\\

It follows from (\ref{07}) that (\ref{05}) admits the GCS
(\ref{06}) if and only if
\begin{eqnarray}
D_t{\eta}=0. \label{08}
\end{eqnarray}
where $D_t$ denotes the total derivative in $t$. Moreover, if
$\eta$ does not depend on time $t$ explicitly, then
\begin{eqnarray}
\eta\prime{E}|_{{L}\bigcap{W}}=0, \nonumber
\end{eqnarray}
where
\begin{eqnarray}
 \eta\prime{(u)E}=\lim_{\epsilon \to
0}\frac{d}{d\epsilon}\eta(u+\epsilon E)\nonumber
\end{eqnarray}
 denotes the
Fr\'echet derivative of $\eta $ along the direction $E$.

The outline of this paper is as follows. In Section 2, we will
classify Eq. (\ref{04}) which admit derivative-dependent
functional separable solutions. Some exact solutions to the
resulting equations are presented in Section 3. Section 4 is a
summary and discussion.

\section{Equations with DDFSSs}
In \cite{Zh1}, we have proved the following theorem:\\
{\bf Theorem 1.} Eq. (\ref{05}) possesses the additive separable
solution
\begin{eqnarray}
u=a(x)+b(t), \nonumber
\end{eqnarray}
if and only if it admits the GCS
\begin{eqnarray}
V=u_{xt}\frac{\partial}{\partial{u}}. \label{09}
\end{eqnarray}
{\bf Theorem 2.} Eq. (\ref{05}) possesses the derivative-dependent
functional separable solution (\ref{03}) if and only if it admits
the GCS
\begin{eqnarray}
V=\frac{(f_{uu}u_t+f_{uu_x}u_{xt})u_x+(f_{uu_x}u_t+f_{u_xu_x}u_{xt})u_{xx}
+f_{u_x}u_{xxt}+f_uu_{xt}}{f_u}\frac{\partial}{\partial{u}},
\label{10}
\end{eqnarray}
{\bf Proof:} Let $v=f(u,u_x)=a(x)+b(t)$, in Theorem 1, after
replacing $u$ by $v=f(u,u_x)$, and simplifying (\ref{09}), we get
\begin{eqnarray}
v=f(u,u_x)=a(x)+b(t)\nonumber
\end{eqnarray}
 if and only if
\begin{eqnarray}
V=v_{xt}\frac{\partial}{\partial{v}}=\frac{[(f_{uu}u_t+f_{uu_x}u_{xt})u_x+(f_{uu_x}u_t+f_{u_xu_x}u_{xt})u_{xx}
+f_{u_x}u_{xxt}+f_uu_{xt}]}{f_u}\frac{\partial}{\partial{u}}\label{vect}.
\end{eqnarray}
And then the assertion holds.

From Theorem 2, we know that equation (\ref{04}) admits the DDFSSs
(\ref{03}) if and only if it admits the GCS \eqref{vect}.

By using the Leibnitz rule on $n$-th order differentiation of
product functions,
we arrive at the following lemma:\\
{\bf Lemma 1. } Assume $G(r)\neq 0$, $F(r)$ and $G(r)$ are
arbitrary smooth functions, then $D_r^iF(r)=0$, $i=0,\ 1,\ ...,\
N$, if and only if $D_r^i\left(\frac{F(r)}{G(r)}\right)=0, \ \
i=0,\ 1,\ ...,\ N$.

 In order to cover the special case $f_u=0$ in (\ref{03}), on the basis of Lemma 1,
 we can take away the denominator $f_u$ in (\ref{10}) and choose $\eta$ and $V$ as the following form,
\begin{eqnarray}
V=[(f_{uu}u_t+f_{uu_x}u_{xt})u_x+(f_{uu_x}u_t+f_{u_xu_x}u_{xt})u_{xx}
+f_{u_x}u_{xxt}+f_uu_{xt}]\frac{\partial}{\partial{u}}. \label{11}
\end{eqnarray}
The invariant condition for (\ref{11}) reads
\begin{eqnarray}
&&V^{(2)}(u_t-A(u,u_x)u_{xx}-B(u,u_x))\nonumber\\
&&=D_t\eta-(A_u\eta+A_{u_x}D_x\eta)u_{xx}-A(u,u_x)D^{2}_x\eta-(B_u\eta+B_{u_x}D_x\eta)\nonumber \\
&&=D_t\eta=0,\label{Dt}
\end{eqnarray}
whenever $D^{i}_x\eta=0$, ($i=0,1,2,\cdots$) and \
$u_t=A(u,u_x)u_{xx}+B(u,u_x)$, where
\begin{eqnarray}
\eta \equiv
(f_{uu}u_t+f_{uu_x}u_{xt})u_x+(f_{uu_x}u_t+f_{u_xu_x}u_{xt})u_{xx}
+f_{u_x}u_{xxt}+f_uu_{xt}. \label{12}
\end{eqnarray}
Substituting (\ref{04}) into (\ref{12}) gives
\begin{eqnarray}
\label{13}
 &\eta =&
(u_{xx}f_{uu_x}+u_xf_{uu})(Au_{xx}+B)+(u_{xx}f_{u_xu_x}+u_xf_{uu_x}+f_u)(Au_{xx}+B)_x
\nonumber\\&\  &+f_{u_x}(Au_{xx}+B)_{xx}\nonumber\\
&=&f_{u_x}Au_{{{\it xxxx}}}+ (f_{u_x}A_{u_xu_x}+f_{u_xu_x}A_{u_x}
 ) {u_{xx}}^{3}+ (  ( f_{uu_x}A_{u_x}+f_{u_xu_x}
A_{u}+2f_{u_x}A_{uu_x} ) u_x\nonumber\\& &
+f_{u_x}A_{u}+f_{uu_x}A+f_{u}A_{u_x}+f_{u_x}B_{u_xu_x}+f_{u_xu_x}B_{u_x}
) {u_{xx}}^ {2}+ (
(f_{uu_x}B_{u_x}+2f_{u_x}B_{uu_x}\nonumber\\&&+f_{u}A_{u}+
f_{{uu}}A+f_{u_xu_x}B_{u} ) u_x+ ( f_{u_x}A_{{uu}}+f_{uu_x}A_{u} )
{u_x}^{2}\nonumber\\&&+ ( 3f_{u_x}A_{u_x}+f_{u_xu_x}A
 ) u_{xxx}+f_{u_x}B_{u}+f_{u}B_{u_x}+f_{uu_x}B
 ) u_{xx}\nonumber\\&&+ ( f_{u}A+f_{u_x}B_{u_x}+ (
 f_{{uu_x}}A+2f_{u_x}A_{u} ) u_x ) u_{xxx}\nonumber\\&&+ (
f_{uu_x}B_{u}+f_{u_x}B_{uu} ) {u_x}^{2}+ ( f_{uu} B+f_{u}B_{u} )
u_x=0.
\end{eqnarray}

In order to determine all the possible $f,\ A$ and $B$ from
\eqref{Dt}, a straightforward substitution leads to
\begin{eqnarray}
\label{15} &D_t\eta=&\frac{\partial}{\partial{t}}[f_{u_x}Au_{{{\it
xxxx}}}+ (f_{u_x}A_{u_x})_{u_x} {u_{xx}}^{3}+ (  (
f_{uu_x}A_{u_x}+f_{u_xu_x} A_{u}+2f_{u_x}A_{uu_x} )
u_x\nonumber\\&&+f_{u_x}A_{u}+f_{uu_x}A+f_{u}A_{u_x}+f_{u_x}B_{u_xu_x}+f_{u_xu_x}B_{u_x}
) {u_{xx}}^ {2}+ (
(f_{uu_x}B_{u_x}+2f_{u_x}B_{uu_x}\nonumber\\&&+f_{u}A_{u}+
f_{uu}A+f_{u_xu_x}B_{u} ) u_x+ ( f_{u_x}A_{uu}+f_{uu_x}A_{u} )
{u_x}^{2}\nonumber\\&&+ ( 3f_{u_x}A_{u_x}+f_{u_xu_x}A
 ) u_{xxx}+f_{u_x}B_{u}+f_{u}B_{u_x}+f_{uu_x}B
 ) u_{xx}\nonumber\\&&+ ( f_{u}A+f_{u_x}B_{u_x}+ (
 f_{{uu_x}}A+2f_{u_x}A_{u} ) u_x ) u_{xxx}\nonumber\\&&+ (
f_{uu_x}B_{u}+f_{u_x}B_{uu} ) {u_x}^{2}+ ( f_{uu} B+f_{u}B_{u} )
u_x ].
\end{eqnarray}
Using the integrable condition between (\ref{04}) and $\eta=0$, we
can express $u_{xxxxxx}$,\ $u_{xxxxx}$,\and \ $u_{xxxx}$ in terms
of $u$, $u_x$, $u_{xx}$ and $u_{xxx}$, and substituting these
expressions into (\ref{15}), we have
\begin{eqnarray}
D_t\eta=&(h_1u_{xx}+h_2)u^{2}_{xxx}+(h_3u^{3}_{xx}+h_4u^{2}_{xx}+h_5u_{xx}+h_6)u_{xxx}\nonumber\\
&+h_7u^{5}_{xx}+h_8u^{4}_{xx}+h_9u^{3}_{xx}+h_{10}u^{2}_{xx}+h_{11}u_{xx}+h_{12}=0,
\label{16}
\end{eqnarray}
or equivalently
\begin{eqnarray}
\label{17}
 h_i=h_i(u,u_x)=0,\ i=1,2,\cdots,12.
\end{eqnarray}
where the expressions for $h_i$ are complicated, and are given in
the appendix A. Eq.(\ref{04}) possesses the DDFSSs (\ref{03}) if
only (\ref{16}) holds, or equivalently, the system of PDEs
(\ref{17}) holds. From the system (\ref{17}), one obtains the
following relations among $A,\ B$ and $f$:
\begin{eqnarray}
0&=&\left( 3+5\sqrt{1+g_0(u) } \right)\ln A-\sqrt{1+g_0( u
)}\ln[-3{A_{u_x}}^{2}g_1(u)( 2+g_0(u) )-g_2 (u)\nonumber\\&&-2g_0(
u ){A}^{3}+ 2\sqrt {3{A_{u_x}}^{2}g_1(u)( 1+g_0(u) ) +2\sqrt
{3}g_0 (u) {A}^{3}}\sqrt {g_1 (u) }A_{u_x} ]\label{18}\\&&
 -2\ln [ A_{u_x} \sqrt {3}\sqrt {g_1 (u)(
1+g_0(u ))}+\sqrt {3{A_{u_x}}^{2}g_1(u)( 1+g_0(u) ) +2g_0 (u)
{A}^{3}} ]\nonumber ,
\end{eqnarray}
\begin{eqnarray}
f_{u_x}&=&f_0 (u) \frac{A_{u_x}}A{\exp\left({-\frac2{3g_1
(u)}\,\int \!{\frac {{A}^{2 }}{A_{u_x}}}{du_x}}\right)},
\label{19}
\end{eqnarray}
\begin{eqnarray}
B&=&\int \! \left\{ \int \!\left[
\left(-4f_{u_xu_x}f_{uu_x}u_x{A}^{2}-2f_{u_xu_x}f_{u}{A}^{2}-2(f_{{u_{{x}
}}})^{2}A_{u}A+2(f_{u_x})^{2}A_{u}A_{u_x}u_x\right.\right.\right.\nonumber\\&&\left.+3
f_{u_x}Af_{u}A_{u_x}-2(f_{u_x})^{2}Au_xA_{
uu_x}+3f_{u_x}f_{uu_x}Au_xA_{u_x}+2
f_{u_x}f_{uu_x}{A}^{2}\right.\label{20}\\&&\left.\left.\left.+2u_xf_{{uu_xu_x}
}f_{u_x}{A}^{2}\right)\frac{1}{{A}^{2}{f_{u_x}}^{2}}\right]{du_x}+b_0(u)
\right \} A{du_x}+b_1(u)\nonumber,
\end{eqnarray}
where  $f_0(u)$, $g_0(u)$, $g_1(u)$, $g_2(u)$, $b_0(u)$ and
$b_1(u)$ are arbitrary functions of $u$.

It seems that it is impossible to obtain the general solution
$A(u,u_x)$ from the transcendental equation (\ref{18}) for
arbitrary $g_0(u),\ g_1(u)$ and $g_2(u)$. It is clear that in
order to find explicit solution of $A(u,\ u_x)$ from (\ref{18}),
the only possible cases are, (i) the factor $A_{u_x}$ appears only
in one of two logarithmic functions and (ii) the ratio of the
coefficients of the two logarithmic functions of (\ref{18}) are
integers. After a lengthy computation and tedious analysis, we finally attain the following results:\\
{\bf Theorem 3.} The equation
\begin{eqnarray*}
u_t=A(u,u_x)u_{xx}+B(u,u_x)
\end{eqnarray*}
admits nontrivial DDFSSs of the form (\ref{03}) with
$f_{u_x}(u,u_x)\not=0$, if it is locally equivalent to one
of the following equations, up to equivalence under translation and dilatation of $u$:\\
(1)
\begin{equation}
u_t=\exp[c_3\phi+\phi_u u_x][u_{xx}+\frac{u_{xx}}{u_x}u_x^2+c_3
u_x+c_1 \phi_u^{-1}]+c_2\phi_u^{-1} ,\label{eq1}
\end{equation}
\begin{equation}
 f \left( u,u_x \right) =\phi_{u} u_x +c_{3
}\phi  +c_4\label{f1};
 \end{equation}
 (2)
 \begin{eqnarray}
 && u_t=u_{xx}+c_1 u_x +\frac {\phi_{uu}}{\phi_u}u_x^2+(c_4+c_5 \phi }){
\phi_{u}^{-1} ,\label{eq2}\\&&
 f \left( u,u_x \right) =\phi_{u} u_x +c_3 \phi +c_{2
};\label{f2}
 \end{eqnarray}
(3)
 \begin{eqnarray}
 && u_t=\left( c_1 u+c_2 \right)
{(-u_x) }^{\alpha-1}u_{xx}-\frac {
2c_2}{1+\alpha}(-u_x)^{\alpha+1}+c_4 u+c_3
 ,\label{eq3}\\ &&
 f \left( u,u_x \right) =\ln \left( -u_x \right);
 \label{f3} \end{eqnarray}
(4)
 \begin{eqnarray}
 && u_t=(-u_x)^{\alpha-1}u_{xx}+c_2 (-u_x)^\alpha+c_3 u+c_4
 ,\label{eq4}\\&&
 f \left( u,u_x \right) =\ln \left( -u_x \right);\label{f4}
  \end{eqnarray}
  (5)
 \begin{eqnarray}
 && u_t=u_x u_{xx}+c_3 {u _x}^2+c_4 {u}^2+c_2 u+c_1
 ,\label{eq5}\\&&
 f \left( u,u_x \right) =\ln \left( -u_x \right) ;\label{f5}
  \end{eqnarray}
 (6)
 \begin{eqnarray}
 && u_t=\frac{1}{c_3(c_1u+c_2)-u_x}\left[(c_1u+c_2)u_{xx}-2c_1u_x^2+\left((2c_3c_1^2
-\frac{c_1c_4}{c_2})u+2c_3c_1c_2-c_4\right)u_x\right.\nonumber\\&&\qquad
\left. +c_3c_1\left(c_3c_1^2+\frac{c_1c_4}{c_2}\right)
u^2+2c_1c_3(c_4-c_3c_1c_2)u+c_2c_3(c_4-c_3c_1c_2)\right],\label{eq6}\\&&
 f(u,u_x)=\ln\left[ c_3(c_1u+c_2)-u_x\right] ;\label{f6}
  \end{eqnarray}
(7)
 \begin{eqnarray}
 u_t&=&(c_1 u+c_3-u_x)^{-1}[u_{xx}- ( c_4 u+c_2) u_x +c_1c_4 {u}^2 \nonumber\\
 &&+ ( c_1 c_2 -{c_{1
}}^2 +c_3 c_4 ) u+ c_3 (c_2-c_1)] ,\label{eq7}\\&&
 f ( u,u_x ) =\ln ( c_1 u+c_3-u_x );\label{f7}
 \end{eqnarray}
 (8)
 \begin{eqnarray}
&& u_t=(c_1-u_x)^{-1} u_{xx}-c_3 \ln ( c_1-u_x ) +c_4 u+c_2
 ,\label{eq8}\\&&
 f ( u,u_x ) =\ln ( c_1-u_x )
 ;\label{f8}
  \end{eqnarray}
(9)
 \begin{eqnarray}
  u_t&=&[
c_1( c_2 u+c_3 )-u_x]^{ 2}[ u_{xx}+(c_2 u+c_3
 )u_x^2-( 2c_1 c_2^2u^2+4c_1 c_2 c_3 u+c_4\nonumber\\
&& +2c_1 {c_3 }^2)u_x+{c_1 }^2c_2^3u^3+2{c_1 } ^2c_2^2c_3
 u^2+c_1c_2(c_4+3c_1c_3^2-c_1c_2)u+c_1^2c_3^3\nonumber\\
 &&+c_1c_3c_4-c_1^2c_2c_3],\label{eq9}\\&& f(u,u_x )=\ln [c_1 (c_2 u+c_3 )-u_x]
 ;\label{f9}
  \end{eqnarray}
  (10)
 \begin{eqnarray}
&&u_t= ( c_1-u_x ) ^{\alpha}u_{{{\it xx}} }+c_2,\quad \alpha\neq
-1,\alpha\neq -2,c_1 \neq 0 ,\label{eq10}\\&& f ( u,u_x ) =\ln
(c_1-u_x )
 ;\label{f10}
  \end{eqnarray}
(11)
 \begin{eqnarray}
 &&u_t= u_{xx}+c_4 ,\label{eq11}\\
&& f \left( u,u_x \right) =c_2 { {\rm arcsinh}} [\tan ( u_x +c_1
)] +c_3;\label{f11}
  \end{eqnarray}
(12)
 \begin{equation}\label{eq12}
 u_t=\frac{6c_3 ^2}{(\phi_u u_x-c_3\phi-c_3c_4)^2}[-u_{xx}-\frac{\phi_{uu}}{\phi_u}u_x^2+2c_3u_x-c_3^2\phi\phi_u^{-1}
 -c_3^2c_4\phi_u^{-1}],
\end{equation}
\begin{eqnarray}
f ( u,u_x ) &=&\frac{1}{
 ( \phi_{u}u_x -c_3 \phi -c_3 c_4 ) ^2}[c_1\phi_u^2 u_x^2 -2c_1 c_3(\phi+c_4)\phi_u u_x\nonumber\\
 &&+c_1 c_3^2(\phi+c_4)^2+c_2c_3^2]
  ;\label{f12}
  \end{eqnarray}
  (13)
\begin{eqnarray}
u_t=\frac {6
\phi^2}{(u_x-c_3\phi)^2}[-u_{xx}+\frac{6\phi_u+c_4}{6\phi}u_x^2-\frac{c_3c_4}{3}u_x+\frac{1}{6}
c_3^2 c_4\phi] ,\label{eq13}
\end{eqnarray}
\begin{eqnarray}
f( u,u_x ) =\frac { c_1 {u_x }^2-2c_1 c_3\phi u_x +( c_2 +c_1 c_3
^2) \phi^2}{(u_x-c_3\phi)^2} ;\label{f13}
  \end{eqnarray}
(14)
 \begin{eqnarray}
 && u_t=\frac {-3 ( c_4 \phi -c_3)^2}{2(\phi
 u_x-1)^2}[u_{xx}+\frac{\phi_{uu}}{\phi_u}u_x^2]
,\label{eq14} \end{eqnarray}
 \begin{eqnarray}
f ( u,u_x ) &=&\frac{1}{ 4( \phi_{u} u_x-1)^2}[4c_1
\phi_u^2u_x^2-8c_1 \phi_u u_x+c_2 c_4^2
\phi^2\nonumber\\
&&-2c_2c_3c_4\phi+4c_1+c_2c_3^2]
 ;\label{f14}
  \end{eqnarray}
(15)
 \begin{equation}\label{eq15}
 u_t=-\frac {-6\phi^2}{(u_x-c_4\phi)^2-\phi^2}[u_{xx}-
 (\frac{\phi_u}{\phi}+c_3)u_x^2+2c_3c_4 u_x+c_3(1-c_4^2)\phi]
 ,
\end{equation}
 \begin{eqnarray}
&&f ( u,u_x ) = c_1 \ln [\frac { ( u_x -c_4\phi) ^2}{ ( u_x -c_4
\phi)^2-\phi^2}] +c_2
 ;\label{f15}
  \end{eqnarray}
 (16)
 \begin{eqnarray}
u_t=\frac{\sqrt{6}}{3}g_u u_x
[u_{xx}+\frac{g_{uu}}{g}u_x^3+\frac{\sqrt{6}}{2}c_1u_x+\frac{c_5}{3c_2
g_u}(g+c_4),\label{eq16}
 \end{eqnarray}
 \begin{eqnarray}
f \left( u,u_x \right) = c_2 \ln [{\frac {g_{u}u_x}{\sqrt {c_3
}}}]
 ;\label{f16}
  \end{eqnarray}
(17)
 \begin{eqnarray}
 && u_t=\frac{\sqrt {6}}3u_x \phi_{u}u_x
u_{xx}+\frac
{\sqrt{6}}{18(c_2\phi-2c_3)}[6(c_2\phi-2c_3)\phi_{uu}-5c_2\phi_u^2]u_x^3
,\label{eq17}\\
&&f(u,u_x)=c_1 \ln \left(\sqrt{\frac {3c_1}{ {c_2\phi-2c_3}}}
\phi_{u}u_x\right)
 ;\label{f17}
  \end{eqnarray}
(18)
 \begin{eqnarray}
 u_t&=&\left( \sqrt{\frac23}\phi u_x +c_2\right
 ) u_{xx}+\sqrt{\frac23}\phi_{u} {u_{
x}}^3+\frac{c_2}{c_5 } \left( c_4 \phi  -{\frac { ( 4c_3 -3c_5 )
\phi_{u} }{ \phi }} \right) {u_x }^2\nonumber \\&&+\frac{\sqrt
{6}{c_2 }^2}{c_5 } \left(c_4 -{\frac { 2(2c_3 -c_5
 ) \phi_{u}  }{ \phi^2}} \right) u_x-{\frac {3{c_2 }^3
 ( 2c_3 -c_5 ) \phi_{u} }{c_{{5
}}  \phi^3}}\nonumber\\&&+{\frac {c_2
 ( 3c_4 {c_2 }^2+2c_1 c_5 ) }{2c_5
}}\phi^{-1},\label{eq18}
\end{eqnarray}
\begin{eqnarray}
f ( u,u_x ) =\ln (\sqrt {6}\phi u_x+3c_2 )
 ;\label{f18}
  \end{eqnarray}
  where $\phi(u)$ and $g(u)$ are arbitrary functions of $u$, and $\alpha_i$, $\mu$, $c_i$, $i+1,2,\cdots,$ are arbitrary
  constants.

\section{Explicit exact DDFSSs}

In this section, we deduce exact solutions of the equations
obtained in the last section by means of the DDFSS ansatz
(\ref{04}). We affirm that the resulting equations enjoy abundant
exact solutions due to their inclusive arbitrary functions and
constants. Now we just give some of which resulting from the
derivative-dependent
functional separable procedure for all the models listed in the theorem 3.\\
{\bf Example 1.} For the equation \eqref{eq1}, to obtain exact
solutions via derivative-dependent functional separable procedure,
one solves the DDFSS ansatz \eqref{03} with \eqref{f1} first and
then substitute the result to the original equation \eqref{eq1} to
fix the concrete functions $a(x)$,\ $b(t)$ and the integration
function. Finally, we find that \eqref{eq1} has an implicit
separable solution
\begin{eqnarray*}
\phi (u)&=&tc_2 +{\frac {c_1 }{{c_3 }^2 } }+{\frac {\ln ( c_2 c_3
) -\ln ( 1- \lambda{\exp\left[c_2 c_3 ( t+a_2 ) \right]}
 ) }{c_3}}+a_2 c_2 -{\frac {c_1 x}{c_3 }}\\&& + \left[ a_1-{\frac {{\exp({c_3 x})} \left(c_3 x-1+\ln
 ( c_3 c_1 ( c_1 -c_3 )
 ) \right) }{c_3 }}\right.\\&& \left.-{\frac {\mu\ln \left(
\lambda{\exp[c_2 c_3 ( t+a_2 ) ]} -1 \right) }{c_3 \lambda}}
\right] {\exp({-c_3 x})}
\end{eqnarray*}
for $c_3\neq0$, where and hereafter $\lambda$, $\mu$, $a_i$,
$c_i$, $i\in Z$ are arbitrary constants.

For $c_3=0$, the equation \eqref{eq1} has two derivative-dependent
functional separable solutions given implicitly by
\begin{eqnarray*}
\phi(u)&=&\frac{1}{2c_1}\,\ln ({\frac {c_1}{1-{e^{c_1\left (x
+a_1\right )}}\lambda}})\ln [{\frac {c_1\left (1-{e^{c_1\left (
x+a_{{1}}\right )}}\lambda\right )}{{e^{2\,c_1\left
(x+a_{{1}}\right )}}{\lambda}^{2}}}]\\&&-\frac12 c_1 x^2-\frac12
\frac {\left (2\,a_{{1}}{c_{{1
}}}^{2}-2\,a_{{0}}c_1+2\,c_1c_4\right )}{c_1}x-\left (\lambda
e^{c_1(-c_4+a_0 )}-c_2\right )t\\&& -\frac{1}{c_1}\left({\rm
dilog}( \frac {e^{c_1 (x+a_1)}\lambda}{e^{c_1
(x+a_1)}\lambda-1})-a_2c_1\right),
 \end{eqnarray*}
and
 \begin{eqnarray*}
\phi(u)=\left(\int \!\ln {\frac {\left (\mu\,c_1 x
-\mu-c_1\lambda\right ){e^{c_1x}}-{
e^{a_{{1}}}}}{c_1^{2}{e^{c_{{1} }x}}}}{dx}-x\ln [\mu \left
(t+a_{{3}}\right )]+c_2 t+a_2\right),
\end{eqnarray*}
where the function ${\rm dilog}(x)$ is the usual dilogarithm
function defined by:
$${\rm dilog}(x)=\int_1^x\frac{\ln(t)}{1-t}{\rm dx}. $$
{\bf Example 2.}\ In the same way as for the last example, we can
find that the equation \eqref{eq2} has the implicit separable
solution
\begin{eqnarray}
\phi (u)&=&\frac{1}{-2c_3 c_5 ( c_5 +{c_3 }^2-c_1 c_3
 ) }\left.\{ \left[ 2 ( c_4 c_{{3 }}{\exp(c_3 x)}-a_1 c_5
c_3 {\exp(( c_5 +{c_3 }^{2 }-c_1 c_3 ) t)}\right.\right.\nonumber \\
&& +a_2 c_5 ( c_5 +{c_3 }^2-c_1 c_3 ){\exp(c_3 x+c_5 t)}
 )  -c_3
a_4c_5 ( 2c_3-c_1 -\sqrt {{c_1 }^2-4c_5 } ) \nonumber \\
&& \times {\exp\left(\frac12 ( 2c_3 -c_1 +\sqrt {{c_1 }^2 -4c_5 }
)x \right)} -a_3 c_3 c_5(2c_3 -c_{1 }+\sqrt {{c_1 }^2-4c_5 } )
\nonumber \\ && \left.\left. \times {\exp\left(-\frac12(c_1 -2c_{{
3}} +\sqrt {{c_1 }^2-4c_5})x\right)}\right] {e^{-c
_{3}x}}\right\}\label{kink}
\end{eqnarray}
for $c_5\neq 0$.

If $c_5=0$, the equation has the DDFSS
\begin{eqnarray*}
\phi(u)&=&-{\frac {c_{{4}}x}{c_{{1}}}}-{\frac
{a_{{1}}{e^{-c_{{1}}x}}}{c_{{1}} \left (c_{{3}}-c_{{1}}\right
)}}+{\frac {c_{{3} }[c_{{4}}+ \mu+c_{{1}}\left
(\mu\,t-c_{{2}}+a_{{2}}+a_{{4}}\right )]-\mu}{{c_{{3}}}^{2}c_{{1}}}}\\
&&+a_{{3}}{\exp[-c_{{3}}\left (x+(c_1-c_3)t\right )]}.
 \end{eqnarray*}

Notice that for $c_1=0,\ c_4=0,\ c_5=0,\ \phi(u) =e^u$, the
equation turns into the potential Burgers equation
\begin{eqnarray*}
u_t=u_{xx}+{u_x}^{2}.
\end{eqnarray*}
It is related to the usual Burgers equation
\begin{eqnarray*}
v_t=v_{xx}+2v{v_x}
\end{eqnarray*}
by $v=u_x$.

In this case, the DDFSS (\ref{kink}) becomes
\begin{eqnarray*}
v=u_x=\{\ln[(a_4+a_2+a_3)c_3^{-1}+a_1\exp(c_3^2t-c_3x)]\}_x.
\end{eqnarray*}

Usually, if the inverse function of $\phi(u)$ in (\ref{eq2}) is
well-defined then the solution (\ref{kink}) denotes a multiple
soliton resonant solution. For instance, setting
$$ \phi(u)=\tan u,$$ then the nonlinear diffusion equation
(\ref{eq2}) becomes
\begin{eqnarray}\label{ex2}
u_t=u_{xx}+c_1u_x+2u_x^2\tan(u)+c_4\cos^2(u)+\frac12c_5\sin(2u)
\end{eqnarray}
and the corresponding solution (\ref{kink}) becomes
\begin{eqnarray}
u&=&\arctan \left\{\frac{2a_3\exp\left[-\frac12\left(\sqrt {
c_1^2-4c_5}+c_1\right)x\right]}{2c_3-c_1-\sqrt{c_1^2-4c_5}}
+\frac{2a_4\exp\left[\frac12\left(\sqrt{c_1^2-4c_5}-c_1\right)x\right]}{2c_3-c_1+\sqrt{c_1^2-4c_5}}\right.\nonumber\\
&&
\left.+\frac{a_2}{c_3}\exp(c_5t)-\frac{c_4}{c_5}+a_1\exp[(c_3^2-c_3c_1+c_5)t-c_3x]\right\}.\label{kink1}
\end{eqnarray}
When $a_2=a_3=a_4=0,\ a_1\neq0$ the solution \eqref{kink1} denotes
a travelling kink solution. \eqref{kink1} is a static kink
solution for $a_2=a_1=a_4=0,\ a_3\neq0$ or $a_2=a_3=a_1=0,\
a_4\neq0$. \eqref{kink1} becomes a instanton solution for
$a_1=a_3=a_4=0,\ a_2\neq0$. Generally, the solution (\ref{kink1})
is a resonant solution of the travelling kink, static kink and the
instanton excitations. Fig.1 is the evolution plot of the single
kink solution \eqref{kink1} with
\begin{equation}
a_1=1,\ a_2=0,\ a_3=0,\ a_4=0,\ c_4=4,\ c_5=4,\ c_1=5,\
c_3=3.\label{k1}
\end{equation}
Fig. 2 is the evolution plot of the resonant solution of the
travelling kink and the static kink while the corresponding
parameters are taken as follows
\begin{equation}
a_1=1,\ a_2=0,\ a_3=1,\ a_4=0,\ c_4=4,\ c_5=4,\ c_1=5,\
c_3=3.\label{k2}
\end{equation}
The resonant solution shown by Fig. 2 denotes the fusion
interaction between kink and anti-kink. Before the interaction,
there are one large travelling kink and one small travelling kink.
After the interaction, the large kink and the small anti-kink
degenerate to a single smaller static kink. The soliton fusion and
fission phenomena can be found in many (1+1)-dimensional
integrable models \cite{ying} and have been observed in some real
physical systems.

\input epsf
\begin{figure}
\epsfxsize=7cm\epsfysize=6cm \epsfbox{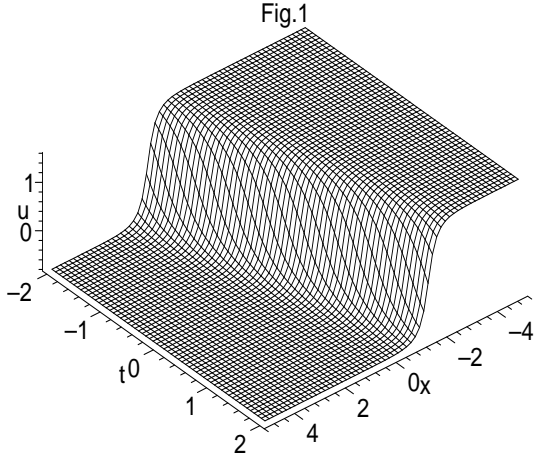}
\caption{Evolution plot of the single kink solution \eqref{kink}
with \eqref{k1}}
\end{figure}

\input epsf
\begin{figure}
\epsfxsize=7cm\epsfysize=6cm \epsfbox{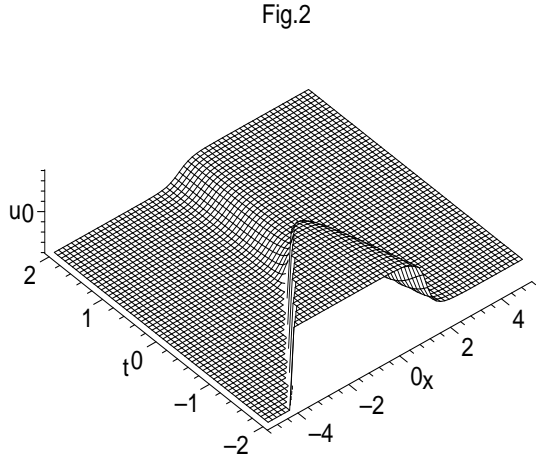} \caption{Kink
fusion interaction expressed by \eqref{kink} with \eqref{k2}.}
\end{figure}

{\bf Example 3.} The equation \eqref{eq3} possesses an explicit
separable solution
\begin{eqnarray*}
&&u= \left[ \int \! \left[ -\gamma\, (c_1 c_3 -c_2 c_4+2\, ( -1)
^{1+2\,\alpha}c_2 c_3 ) {{\exp(a_{{2 }}\alpha\,c_4 )+( \alpha-1 )
c_4 t}}-c_3 {\exp(-c_4t)} \right] \right. \\&&\qquad \left.
\times\left[ ( -c_1 +2c_{2 } ( -1 ) ^{2\,\alpha} )\gamma {\exp(c_4
\alpha\, ( t+ a_2 ) )}-1 \right] ^{-{\frac {c_1 -c_1 \alpha+2\,
\alpha\,c_2 ( -1 ) ^{2\,\alpha}}{\alpha\, ( -c_{{1 }}+2\,c_2 ( -1
) ^{2\,\alpha} ) }}}{dt}+a_3 \right ]\\&&\qquad \times \left[ (
-c_1 +2c_2 ( -1
 ) ^{2\,\alpha} ) \gamma{\exp(c_4 \alpha\, ( t+a_2
 ))}-1 \right] ^{{\frac {c_1 }{\alpha\, ( -c_1 +2\,c
_2 ( -1 ) ^{2\,\alpha} ) }}}{\exp(c_4 t)}\\&&\qquad -\frac{{
\gamma}^{\frac1{\alpha}} ( \alpha\, ( x+a_1 )
 ) ^{{\frac {1+\alpha}{\alpha}}}}{1+\alpha} \left[ {\frac {c_4{\exp(\alpha_4c_4(t+a_2))} }{
 ( c_1 +2\, ( -1 ) ^{1+2\,\alpha}c_
2 )\gamma {\exp(c_4 \alpha\, ( t+a_2 ) )}+1}}
 \right] ^{\frac1{\alpha}}.
\end{eqnarray*}

If $c_1=0,c_2= (-1) ^{n},c_3=0,\alpha=n $, the above equation
becomes
\begin{eqnarray*}
u_t=-{u_x}^{n-1}u_{xx}+{\frac {2}{1+n }}u_x^{n+1}+c_4u
\end{eqnarray*}
which is just the equation $(A.2)$ of Theorem 2 in \cite{Zh2} for
 $$g_1=0,b_1=-1,b_2=0,\beta=-2,\gamma=-c_4.$$
Here we obtain its new variable separation solution, the DDFSS,
which is given explicitly by
\begin{eqnarray*}
&u=& \left\{ \int \!{\frac {\gamma\,c_4 {\exp[c_4(a_{ {2}}n+t
\left( n-1 \right))]}}{2\,\gamma\, {\exp[c_4n \left( t+a_2
 \right)]}-(-1) ^{n}}}{dt}+a_3 \right\}
 {e^{c_4 t}}\\&&-\frac{1}{1+n}
 [ n \left( x+a_1 \right)  ]^{\frac {1+n}{n}}
 \left\{ {\frac{2\, (-1)^{n+1}\gamma\,
 {\exp[c_4n \left( t+a_2 \right) ]}+1}{c_4\gamma}}
 \right\}^{\frac{1}{n}}{\exp(c_4 t+a_2c_4)}.
\end{eqnarray*}
{\bf Example 4.}\ An explicit separable solution of \eqref{eq4}
reads
\begin{eqnarray*}
&&u= -\left(\int \!{\exp(a ( x )+c_3(a_3+t))}{dx}
 -\frac{\mu \exp(c_3t)}{\lambda}\right)
 \left( {\frac {c_3 \alpha}{1-\lambda\,{\exp(c_3 ( \alpha-
1 ) ( t+a_3 ) )}}} \right) ^{{\frac 1{ \alpha-1}}}\\ &&\qquad
+a_2{\exp(c_3 t )} -\frac{c_4}{c_3},
\end{eqnarray*}
for $c_3\neq 0$, where $a(x)$ satisfies
$$a''( x ) +\alpha a'^2+c_2
a'-\frac{\lambda}{\alpha}e^{(1-\alpha)a(x)}=0.$$

If $c_3=0$, an DDFSS is given by
\begin{eqnarray*}
u=-{e^{a_{{0}}}}\int \left (\frac{1}{c_2}
(a_{{1}}{c_{{1}}}^{2/3}{e^{{ \frac
{\alpha\,a_{{0}}{c_1}^{2/3}+\alpha c_2 e^{-\alpha a_0}x}{
{c_{{1}}}^{2/3}}}}}+\alpha a_2c_2 )\right )^{{
\frac{1}{\alpha}}}{dx}+\left( c_4+\alpha\sqrt[3]{c_1}a_2c_2\right
)t+a_{{3}}.
 \end{eqnarray*}
\\
{\bf Example 5.}\ The equation \eqref{eq5} has an explicit
separable solution
$$u=-{e^{b (t) }}\int \!{e^{a (x) }}{dx}+s
 (t),$$
 for $c_4\neq 0,$
where $a(x), \ b(t)$ and $s(t)$ satisfy
\begin{eqnarray*}
&&a''' +(5 a'(x)+2c_3) a''+2  a'^3+c_3 a'^2 +2c_4=0,\\
&&b (t) =\ln
 \frac{4a_1 ({c_5 }^2-4c_1 c_4
 ) }{
 4( {c_5 }^2-4c_1 c_4 ) ({\mu}^2+4\lambda c_4)-{a_1 }^2
 ( {e^{-\sqrt {c_5^2-4c_1 c_4}t}} +a_2 \sqrt {c_5^2-4c_1 c_4 })^2}\\
 && \qquad -\sqrt {{c_5 }^2-2c_1 c_4 }t,\\
&&s (t) ={\frac {-c_5 +b' ( t
 ) +\mu{e^{b (t) }}}{2c_4 }}.
\end{eqnarray*}

If $c_4=0$, the DDFSS of \eqref{eq5} becomes
\begin{eqnarray*}
u=\frac{\left(-\lambda^2c_1-\mu c_5^2-c_5^2\lambda\int e^{ a
\left( x \right) }{\rm dx} \right)e^{c_5
 \left(t+a_0 \right) }+\lambda\left(- a_1c_5\left({e^{c_5t}}+c_1+\lambda\,{e^{c_5 \left(2\,t+a_0 \right) }}
\right)\right)}{ c_5\lambda\left(\lambda\,{e^{c_5 \left( t+a_0
 \right) }} -1\right) },
 \end{eqnarray*}
where $a(x)$ satisfies
 \begin{eqnarray*}
a''-2a'^2+c_3 a'-\lambda e^{-a(x)}=0.
\end{eqnarray*}
\\
{\bf Example 6.} An explicit variable separable solution of Eq.
(\ref{eq6}) has  the form
\begin{eqnarray*}
 &&u=\frac{1}{{\mu}{c_1 }}\left[ 2\,{c_1 }^2{\exp\left({-\frac { \left( c_1
\mu c_2-2 c_1^2c _4 \right) t+c_2 \mu x-4\,c_2 a_1 c_1}{2c_1 c_2
}}\right)}-c_2\mu{\exp\left(-c_3 c_1
x\right)}\right.\\&&\qquad\left. +\frac { a_ 2\mu\,c_1( \mu\,c_2
+2\,c_1 c_4 )}{2c_2 }t \right] {e^{c_1 c_3 x}}.
\end{eqnarray*}

If $c_1=0,$ the DDFSS of \eqref{eq6} is given explicitly by
\begin{eqnarray*}
u=-\sqrt {c_{{2}}a_{{1}}} \left( t+a_{{3}} \right) \tan \left(
\frac12\,{ \frac {\sqrt {c_{{2}}a_{{1}}} \left( x+a_{{2}} \right)
}{c_{{2}}}}
 \right) +c_4 t+c_2c_3 x +a_4+\frac12 a_1a_2a_3.
 \end{eqnarray*}

If $c_4=0,c_3=0 $, the equation \eqref{eq6} becomes
\begin{eqnarray*}
u_t=-{\frac { \left( c_1u+c_2 \right) u_{xx}}{u_{{ x}}}}+2\,c_1u_x
\end{eqnarray*}
which is equivalent to the equation $(A.3)$ of Theorem 2 in
\cite{Zh2} for
 $$ g_1=0,\beta=0,n=0,b_1=-c_2,b_2=-c_1,g_2=0. $$
It has a new solution
\begin{eqnarray*}
u= \frac{1}{{c_1}{\mu}}\left[ 2\,{c_1}^{2}{\exp\left(-{\frac
{c_1\mu\,t+x\mu-4\,a_{{ 1}}c_1}{2c_1}}\right)}-\mu\,c_2+a_2\mu
c_1{\exp\left(\frac12\,\mu\,t\right)} \right].
\end{eqnarray*}\\
{\bf Example 7.}\ The equation \eqref{eq7} possesses an explicit
variable separation solution
\begin{eqnarray*}
&&u={\frac { ( \mu\,c_1 -c_1 {e^{c_4 ( t+a _2 ) }} ) \int
\!{e^{-c_1 x+a (x) }}{d x}+a_1 c_1 c_4 e^{c_4 t}-c_3c_4e^{-c_1 x}
+ \lambda c_1 }{c_1 c_4 e^{-c_1 x} }}
 \end{eqnarray*}
 for $c_1c_4\neq0$,
 {\rm where} $ a(x)$ {\rm satisfies}
 \begin{eqnarray*}
&&a''(x) - c_1 a'(x) -\mu\,{e^{a
 (x) }}-c_3 c_4 -c_1^2+c_1c_2=0.
\end{eqnarray*}

If $c_1=0,\ c_4=0$, its DDFSS is given explicitly by
\begin{eqnarray*}
u={\frac { \left( {e^{c_4 \left( t+a_1 \right) }}-\lambda
 \right) \int {e^{a \left( x \right) }}{dx}}{c_4}}+c_3x+a_2e^{c_4 t}-\frac {c_2
+\mu}{c_4} ,
 \end{eqnarray*}
where $a(x)$ satisfies
 \begin{eqnarray*}
\pm \int _{\mbox {{\tt }}}^{a \left( x \right) }\!{\frac {1}{
\sqrt { -2\lambda {e^{s}}+2c_4c_3s+a _3  }}}{ds}=x+a_4.
\end{eqnarray*}

If $c_1\neq0$, $c_4=0$,  its DDFSS is given explicitly by
\begin{eqnarray*}
u=\lambda \left( t+a_1 \right) {e^{c_1x}}\int {e^{a(x)-c_1
x}}{dx}+ (\mu\,t+ a_2) {e^{c_1x}}-\frac {c_3}{c_1} ,
 \end{eqnarray*}
where $a(x)$ satisfies
 \begin{eqnarray*}
a''-c_1a'+\lambda e^{a(x)}+c_1c_2-c_1^2=0.
\end{eqnarray*}

If both $c_1$ and $c_4$ are zero, the derivative-dependent
functional separable solution of \eqref{eq7} should be
 \begin{eqnarray*}
u=\frac{b_1(c_3x+b_2t+b_0)\exp[b_3(x+b_4)]+c_3b_1x+b_1(b_2-2b_3c_2)t-2b_3c_2t_0+b_0b_1}{b_1[\exp[(x+b_4)b_3]+1]},
\end{eqnarray*}
with $b_0,\ b_2,\ b_3,\ b_4$ and $t_0$ being all arbitrary
constants.
\\{\bf Example 8.} For $c_4\neq 0$, the derivative-dependent
functional separable solution of Eq. \eqref{eq8} reads
\begin{eqnarray*}
&&u= \frac{1}{c_4\lambda}\left(\exp[c_4 (t+a_3) ]-\lambda\right)
 \left[\lambda\int \exp[a (x)] {dx}-c_3 \ln
 ( -{\frac {\exp[a_3 c_4 +c_4 t]-\lambda}{c_4 }}
 ) \right] +c_1 x\\&&+a_2 {e^{c_4 t}}+{\frac {c_3 {e^{c_4 ( t+a_3 ) }}t }{\lambda}}-{\frac {\mu+c_2
}{c_4 }},
 \end{eqnarray*}
 where $ a(x)$ satisfies
 \begin{eqnarray*}
&&a''(x) +c_3 a'(x)+\lambda e^{a(x)}- c_1 c_4 =0.
\end{eqnarray*}

If $c_4=0$, the DDFSS of \eqref{eq8} is changed to
\begin{eqnarray*}
u&=&-\lambda\, \left( t+a_2 \right) \int \!{e^{a \left( x \right)
}} {dx}-c_3 \left( t+a_2 \right) \ln [ \lambda\,
 \left( t+a_2 \right)  ] \\&&+ \left( \mu+c_3+c_2
 \right) t+c_1x+c_3a_2+a_1 ,
 \end{eqnarray*}
where $a(x)$ satisfies
 \begin{eqnarray*}
a''(x) +c_3a'(x)-\lambda e^{a(x)}=0.
\end{eqnarray*}

{\bf Example 9.}\
$$u= \left[ -{\frac {\sqrt
{{e^{2\,c_2 \left( t+a_2 \right) }}- \lambda} \left(\lambda \int
\!{e^{-c_1 c_2 x+a \left( x \right) }}{dx} +\mu \right) }{\sqrt
{c_2 }\lambda}}+a_1 {e^{c_2 t }}
 \right] {e^{c_1 c_2 x}}-{\frac {c_3 }{c_2 }},$$
with $a(x)$ being a solution of
\begin{eqnarray*}
a''\left( x \right)-  a'(x)^2+ ( c_4-2\,c_1c_2) a' \left( x
\right)  -\lambda e^{2a(x)}+c_1 c_2(c_4 -c_1 c_2 )=0,
\end{eqnarray*}
is an exact solution of Eq. (\ref{eq9}) via the variable
separation formula \eqref{03} with \eqref{f9}.

When  $c_2=0$, the DDFSS is given explicitly by
\begin{eqnarray*}
u=-\sqrt {2}\sqrt {\lambda\, \left( t+a_2 \right) } \left( \int \!
{e^{a \left( x \right) }}{dx}-\mu\,\lambda \right)
+a_1+c_1c_3x+c_3t ,
 \end{eqnarray*}
where $a(x)$ satisfies
 \begin{eqnarray*}
a''(x)-a'(x)^2+c_4 a'(x)-\lambda e^{2 a(x)}=0.
\end{eqnarray*}

{\bf Example 10.}\ The equation \eqref{eq10} has the following
special variable separable solution
$$ u=-\left( {\frac
1{\mu\,\alpha \, \left( t+a_4 \right) }} \right)
^{\frac{1}{\alpha}}\int \!{e^{a \left( x \right) }}{dx} +c_1 x-
 \left( {\frac 1{\mu\,\alpha\, \left( t+a_4 \right) }} \right) ^
{{\frac {\alpha+1}{\alpha}}}\lambda\,\alpha\, \left( t+a_4
 \right) +c_2 t+a_3 ,
 $$
 where $a(x)$ satisfies
 $$\pm  \sqrt {(\alpha+2)}\int _{\mbox {{\tt \ }}}^{a \left( x
 \right) }\!{\frac {{e^{\left( \alpha+1 \right)\xi }}}{\sqrt {a_{{
1}}-2\,\mu\,{e^{ \left( \alpha+2 \right)\xi }}}}}{d\xi}-x-a_2
=0.$$

{\bf Example 11.}\ The equation (\ref{eq11}) is a trivial linear
diffusion equation which allows of course infinitely many product
variable separation solutions. It is easy to see that it allows a
DDFSS which is simply equivalent to a trivial special additive
separation solution
\begin{eqnarray*}
&u= &a_1 x+c _4 t.\hspace{0cm}
\end{eqnarray*}

{\bf Example 12.} The DDFSS of (\ref{eq12}) with $c_2\neq0$ is
given by
\begin{eqnarray*}
\phi (u) &=& {-i\sqrt {c_2 }e^{x}}\left[ \int \!{\frac {  e^{
x}}{\sqrt {-a_3{e^{-2x}} -a_2{e^{2x}} +2\sqrt {a_2 a_3 } \tanh
\left( {\frac { 24\sqrt{a_2a_3}( t+a_1 ) }{c_2 }} \right)
}}}{dx}\right.
\\&&\left.+\frac{1
}{ {a_2 }{a_3 }}\arctan \left( \sqrt {2\sqrt {a_3 a_2 } \tanh
\left( {\frac { 48\sqrt {a_2a_3 }( t+a_1 )}{c_2 }} \right) -a_3
-a_2 } \right)\right]
\\&&-a_4e^{x}-c_4.
\end{eqnarray*}

{\bf Example 13.} The equation (\ref{eq13} has the following
special solution
\begin{eqnarray*}
&&\int _{\mbox {{\tt \ }}}^{u}\! \left( \phi \left( s \right)
\right) ^ {-1}{ds}=-2\,{\frac {\sqrt { \left( 3{\mu}^2t+ +\sqrt
{3}\mu\,c_2x + 3a_1 c_2+3a_2 c_2 -3c_1 c_2 \right) }}{\mu}}+c_3
x+c_4 t.
\end{eqnarray*}

There are three special cases of \eqref{eq13}, which are known in literature:\\
(i) If $c_3=0,\phi (u) =c_0$, \eqref{eq13} becomes
\begin{eqnarray}\label{13a}
u_t=-6\,{\frac {u_{xx}}{{u_x}^{2}}}+c_4,
\end{eqnarray}
which has a DDFSS given explicitly by
\begin{eqnarray*}
u=\left(-2\,{\frac {\sqrt {\sqrt {3}\mu c_2x+3\mu^2
t+3(a_1c_2+a_2c_2-c_1c_2) }}{\mu}}+c_4t\right).
 \end{eqnarray*}
This equation is equivalent to the equation (19) of Example 3.1 in \cite{Zh2} for $ n=-1,\alpha=0 $;\\
(ii)  If $c_3=0,\phi (u) =u $, \eqref{eq13} is simplified to
\begin{eqnarray}\label{13b}
u_t=-6\,{\frac {{u}^{2}u_{xx}}{{u_x}^{2}} }+ \left( 6+c_4 \right)
u ,
\end{eqnarray}
which has the DDFSS
\begin{eqnarray*}
u=\exp\left(-2\,{\frac {\sqrt {\sqrt {3}\mu
c_2x+3\mu^2t+3(a_1c_2+a_2c_2-c_1c_2) }}{\mu}}+c_4t\right).
 \end{eqnarray*}
Eq. \eqref{13b} is equivalent to the equation (26) of Example 3.2
in \cite{Zh2} for $ n=-1,\alpha=0 $, the equation (26) is a
generalization of the curve shortening equation
\begin{eqnarray*}
w_t={w}^{2}w_{xx}+{w}^{3}.
 \end{eqnarray*}
 (iii)  If $c_4=0,c_3=0,\phi (u)
={e^{u}}$, \eqref{eq13} becomes
\begin{eqnarray}\label{13c}
u_t=-6\,{\frac {{e^{2\,u}}u_{xx}}{{u_x}^{ 2}}}+6{e^{2\,u}} ,
\end{eqnarray}
which has the DDFSS
\begin{eqnarray*}
u=\ln  \left( {\frac {\mu}{  2\,\sqrt {\sqrt{3}\mu c_2 x
+3\mu^2t+3(a_1c_2+a_2c_2-c_1c_2)}-a_3\mu
 }} \right).
 \end{eqnarray*}
 After transformation $u=w/2$, \eqref{13c} is transformed to
 \begin{eqnarray*}
w_t=-24\,{\frac {{e^{w}}w_{xx}}{{w_x}^{2} }}+12{e^{w}},
 \end{eqnarray*}\\
  which is equivalent to the equation (23) in the example 3.4 in \cite{Zh2} for
  $ \beta=0,\sigma=-\frac12,n=-1,\alpha=0 $.

{\bf Example 14.}\ The equation \eqref{eq14} admits a special
separable solution given implicitly by
\begin{eqnarray*}
\phi (u) &=& - \frac{1}{{c_4 }^{2}{c_2 }}\left[ 2c_4 \sqrt
{a_0c_2x +3{a_0}^2t-c_{{1 }}c_2 +b_0c_2 }+2a_0-c_3 c_2 c _4\right.
\\&&\left. -a_1{c_4 }^2c_2{\exp\left({\frac {c_4 ( 2\sqrt {a_0c_2x
+3{a_0}^2t-c_1 c_2+b_0c_2 }-3c_4 a_0t ) }{2a_0}}\right)} \right].
\end{eqnarray*}

{\bf Example 15.}\ The DDFSS of \eqref{eq15} has the form
\begin{eqnarray*}
\int _{\mbox {{\tt \ }}}^{u}\! ( \phi ( s ) ) ^
{-1}{ds}&=&\frac{c_4 c_1 }{a_2}\pi i +\frac{c_1 }{a_2}\ln \left[
{\exp\left({\frac {3{a_2 }^2t+a_3 c_1+a_1 c_1 +c_1 c_2 }{{c_{
1}}^2}}\right)}\right.\\&&\left. -2{\exp\left({\frac { c_1( 2a_3
+2a_1 +a_2 x
 )+6{a_2 }^2t}{{c_1 }^2}}\right)}\right. \\ & &\left. -2i
\sqrt {-{\exp\left({\frac { c_1( a_2 x+a_3 +a_1 )  +3 {a_2
}^2t}{{c_1 }^2}}\right)}+{\exp\left({\frac {c_2 }{c_{1
}}}\right)}}\right.\\&&\left.\times {\exp\left({\frac { c_1( 3a_3
+3a_1 +a_2 x ) +9{a_2 }^2t}{2{c_1 }^2}}\right)}\right ] -{\frac {(\ln 2)c_1 }{a_2 }}\\
& &+{\frac {c_4 a _2c_1x + ( -3 {a_2 }^2+6c_3 a_2 c_1
 ) t+i{c_1 }^2\pi + ( a_4a_2 -a_3 -a_1 -c
_2 ) c_1 }{c_1 a_2 }}.
\end{eqnarray*}

{\bf Example 16.}\
\begin{eqnarray*}
g (u) &=&\frac{\lambda}{\mu}\, {\exp\left({\frac
{2\mu\,t+6\,a_2}{3c_2 }}\right)}+ \left[ -3\,c_2 a_1 {e^{{-\frac
{\mu\,t+3\,a_2 }{3c_2 }}}}{\mu}^{-1}\right. \\&& \left.
+\frac{1}{{c_1 }^{3/2}\sqrt {c_2 }\mu}\, \left(- \sqrt 2\mu\,\sqrt
{-{e^{-\sqrt 6c_1 ( x+a_3
 ) }}+\lambda}+\sqrt 2(\ln 2)\mu\,\sqrt {\lambda}
 \right.\right. \\&& \left. \left. +\sqrt 2\mu\,\sqrt {\lambda}\ln ( \lambda+\sqrt {
\lambda}\sqrt {-{e^{-\sqrt 6c_1 ( x+a_3
 ) }}+\lambda} ) +\sqrt {3}\mu\,\sqrt {\lambda}c_1
 ( x+a_3 )\right. \right.\\&& \left.\left.+3\,a_4\sqrt {c_3 }\mu\,{c_1 }^{3/
2}\sqrt {c_2 }\right) \right] {e^{{\frac {\mu\,t+3\,a_2 }{3c_2
}}}}
\end{eqnarray*}
is a special DDFSS of \eqref{eq16}.\\
{\bf Example 17.}\ The equation \eqref{eq17} for $c_2\neq0$ has a
DDFSS determined implicitly by
\begin{eqnarray*}
&\phi (u) =&{\frac {1}{2916{c_2}{c_1}^{3}}}
 \left[ 243{c_2}^{2}{c_1}^{2} {e^{{\frac {2a_2}{c_1}}}}
 \left({e^{{\frac {2a_1}{c_1}}}}{x}^{2}+{a_3}^{2} \right) -18\sqrt
 {6}{c_2}^{3}c_1
{e^{{\frac {4a_2}{c_1}}}}\left( {e^{{ \frac
{4a_1}{c_1}}}}x+a_3{e^{{\frac {3a_1} {c_1}}}}
\right)t\right.\\&&\left. +486{e^{{\frac {2a_2+a_1}{c_{{
1}}}}}}{c_1}^{2}{c_2}^{2}a_3x+2{e^{6{\frac {a_2+a_
1}{c_1}}}}{c_2}^{4}{t}^{2}+5832{c_1}^{3}c_3
 \right].
\end{eqnarray*}

For $c_2=0$, the equation \eqref{eq17} enjoys an exact DDFSS in
the form
\begin{eqnarray*}
\phi \left( u \right) =-\frac{\sqrt{6}}{3c_1\lambda^2(t+a_2)}
\left( -i\lambda{c_{{1}}}^{3/2}\sqrt { c_{{3}}}\int \!{e^{{\frac
{a \left( x \right) }{c_{{1}}}}}}{dx}
-\mu\,c_{{1}}+a_{{1}}{\lambda}^{2} \left( t+a_{{2}} \right)
\right),
 \end{eqnarray*}
where $a(x)$ satisfies
 \begin{eqnarray*}
\sqrt {c_{{3}}}\int _{\mbox {{\tt }}}^{a \left( x \right) }\!{e^{
{\frac {2s}{c_{{1}}}}}}{\frac {1}{\sqrt {-{e^{{\frac
{3a_{{3}}}{c_{{1 }}}}}}+i{c_{{1}}}^{\frac32}\sqrt
{c_{{3}}}\lambda\,{e^{{\frac {3s}{c_1}}}}}}}{ds}=x+a_{{4}}.
\end{eqnarray*}
\\
 {\bf Example 18.} For the equation \eqref{eq18}, a special DDFSS is given by
\begin{eqnarray*}
&&\int _{\mbox {{\tt \ }}}^{u}\!\phi ( s ) {ds}=\frac 1 {\sqrt
{6}} \left[ {e^{b (t) }}\int \!{e^{a (x) }}{dx}-3c_{{2 }}x-s (t)
\right] ,
\end{eqnarray*}
with
\begin{eqnarray*}
\frac {\phi_{u} }{\phi^2}&=&\frac{1}{6c_2 ( 2c_3 -c_5 ) {e^{2a ( x
 ) }}{e^{2b (t) }}}\left[-3\sqrt 6 c_5 b'(t) \int \!{e^{a (x) }}{d
x}{e^{b (t) }}+3\sqrt 6 c_5 s'(t) \right.\\&&\left.+\sqrt 6 c_5
a'(x) {e^{2a ( x
 ) }}{e^{2b (t) }}+3c_4 c_2 {e^{2a
 (x) }}{e^{2b (t) }}+18c_1 c_2 c_{5}\right].
\end{eqnarray*}

Given any $\phi(u)$, then $a(x)$, $b(t)$ and $s(t)$ are determined
by the above two relations, thus the separable solution $u$ is
determined.

\section{Summary and discussion}

In summary, we have brought forward a new conception of DDFSS to
nonlinear evolution equations. Taking the generalized nonlinear
diffusion equation as a concrete example  and using the GCS
approach, we have obtained a complete list of explicit canonical
forms for such equations which admit the DDFSSs. As the
consequence, some exact explicit solutions to the resulting
equations have been obtained via solving the DDFSS ansatz
\eqref{03}. The approach also provides a symmetry group
interpretation to the DDFSSs. Our approach is more general than
the others due to the involvement of the derivative-dependent
functional separable function in the ansatz (\ref{03}).
Subsequently, we can obtain a good many new nonlinear models which
can be solved by means of generalized nonlinear variable
separation procedures. Some new exact solutions of some known
models are given explicitly. Several different types of localized
excitations of some complicated nonlinear diffusion equations have
been found via the DDFSS approach.

Though the variable separation approach has been developed in
several different directions \cite{Zh1}--\cite{Lou1}, it is still
far beyond of perfect. There are some important problems should be
studied further. One of the most important problem may be how to
unify all the known informal variable separation approaches?
Perhaps, we can propose a unified variable separation ansatz in a
most general way ($u=u(x_1,\ x_2,\ ...,\ x_n),\ G_j\equiv
G_j(\xi_1,\ ...,\ \xi_{m_j}),\ \xi_k=\xi_k(x_1,\ ...,\ x_n),\
k=1,\ ...,\ m_j,\ m_j<n$)
\begin{eqnarray}\label{GA}
f(x_1,\ x_2,\ ...,\ u,\ u_{x_i},\ u_{x_ix_j},...)=g(x_1,\ x_2,\
...,\ G_j,\ G_{j\xi_i},\ G_{j\xi_i\xi_k},\ ...).
\end{eqnarray}
Though all the ansatzs of the known informal variable separation
approaches are the special cases of \eqref{GA}, the concrete
realization procedures for different known approaches are quite
different. Can we find a universal method, say, the GCS method, to
realize the generalized variable separation ansatz \eqref{GA}?

\centerline{\bf \large APPENDIX A. Expressions of $h_i$ of
(\ref{17})}


It takes a dozen pages to write down $h_i,\ i=1,\ 2,\ ...,\ 12$,
explicitly in terms of $f,\ A$ and $B$. For simplicity, we display
them by introducing the following notations:
$$
\Gamma_0=(f_{u_x} B_u)_{u} u_{x}^2+(f_{u} B)_{u}
u_{x},\eqno(A.1)$$
$$\Gamma_1=(f_{u_x} A_{u_x})_{u_x},\eqno(A.2)$$
$$\Gamma_2=(f_{u_x} A)_{u}+[(f_{u_x} A_{u_x})_{u}+(f_{u_x} A_{u})_{u_x}] u_{x}
+(f_{u_x} B_{u_x})_{u_x}+f_{u} A_{u_x},\eqno(A.3)$$
$$\Gamma_3=f_{u_xu_x} A+3 f_{u_x} A_{u_x},\eqno(A.4)$$
$$\Gamma_4=(f_{u_x} A_{u})_{u} u_{x}^2+[(f_{u_x} B_{u_x})_{u}+(f_{u_x} B_u)_{u_x}
+(f_{u} A)_{u}] u_{x}+(f_{u} B)_{u_x}+f_{u_x} B_u,\eqno(A.5)$$
$$\Gamma_5=(2 f_{u_x} A_{u}+f_{uu_x} A) u_{x}+f_{u} A+f_{u_x} B_{u_x},\eqno(A.6)$$
$$F_i=\Gamma_i f_{u_x}^{-1} A^{-1}, \ i=0,1,\cdots,5, \eqno(A.7)$$
$$G_1 = F_3 F_1-F_{1u_x},\eqno(A.8)$$
$$ G_2 = -F_{1u} u_{x}-F_{2u_x}+F_5 F_1+F_3 F_2,\eqno(A.9) $$
$$G_3 = -3 F_1+F_3^2-F_{3u_x},\eqno(A.10) $$
$$ G_4 = F_5 F_2-F_{2u} u_{x}+F_3 F_4-F_{4u_x},\eqno(A.11) $$
$$ G_5 = -F_{5u_x}+2 F_3 F_5-F_{3u} u_{x}-2 F_2,\eqno(A.12) $$
$$ G_6 = -F_{4u} u_{x}+F_3 F_0+F_5 F_4-F_{0u_x},\eqno(A.13) $$
$$ G_7 = F_5^2-F_4-F_{5u} u_{x},\eqno(A.14) $$
$$ G_8 = F_5 F_0-F_{0u} u_{x}. \eqno(A.15) $$
With the help of the above notations, $h_i,\ i=1,\ 2,\ ...,\ 12$
in (\ref{17}) read
$$h_1=-9 F_3 A_{u_x}+2 A G_3+15 A_{u_xu_x},\eqno(A.16)  $$
$$h_2=(-2 F_3 A_{u}-F_{3u} A+12 A_{uu_x}) u_{x}-7 F_5 A_{u_x}+4 A_{u}+3 B_{u_xu_x}+F_{5u_x} A+A G_5, \eqno(A.17) $$
$$h_3=(5 G_3+F_{3u_x} -4 F_3^2 -F_1) A_{u_x} -3 F_3 A_{u_xu_x}+10
A_{u_xu_xu_x}$$
$$ +(F_{1u_x} +4 G_1+G_{3u_x}-2 F_1F_3)A,\eqno(A.18)
$$
$$h_4=A G_{5u_x}+F_{5u_x} A_{u_x}-4 F_2 A_{u_x}-F_3^2 B_{u_x}-2 F_3 A_{u}+3 F_1 B_{u_x}
-4 F_5 A_{u_xu_x}$$
$$+4 A_{u} u_{x} G_3+A G_{3u} u_{x}+F_{3u_x} A_{u}
u_{x} +F_{2u_x} A-3 F_1 A F_5-8 F_3 A_{u_x} F_5$$
$$+B_{u_x} G_3-5F_3 u_{x} A_{uu_x} -F_2 A F_3+5 A_{u_x} G_5+6 F_1
A_{u} u_{x} -3 F_3^2 A_{u} u_{x}$$$$+24 u_{x} A_{uu_xu_x}+3 A
G_2+F_{3u} A+F_{3u_x} B_{u_x}+22
A_{uu_x}+6B_{u_xu_xu_x},\eqno(A.19)
$$
$$h_5=(18 A_{uuu_x}-2 F_3 A_{uu}) u_{x}^2+((B_{uu_x}-6 F_5 A_{u}) F_3-7 F_5 A_{uu_x}+4 A_{u} G_5+F_{3u_x} B_u
$$ $$\qquad +A G_{5u}+F_{5u_x} A_{u}+12
B_{uu_xu_x}+4 F_2 A_{u}+16 A_{uu}) u_{x} +(-2 F_5 B_{u_x}+B_u)
F_3$$ $$-7 F_4 A_{u_x}-4 F_5^2 A_{u_x} +(-2 F_2 A-3 A_{u}
-B_{u_xu_x}) F_5+F_{3u} B+10 B_{uu_x}+B_u G_5$$
$$+F_{5u_x} B_{u_x}+F_{5u} A+A
G_{7u_x}+5 A_{u_x} G_7+2 A G_4+2 F_2 B_{u_x}+A F_{4u_x},
\eqno(A.20)$$
$$h_6=4 A_{uuu} u_{x}^3+(F_3 B_{uu}-3 F_5 A_{uu}+6 B_{uuu_x}) u_{x}^2+(4 A_{u} G_7+F_{5u_x} B_u$$
$$-F_5 B_{uu_x} +A G_{7u}+4 B_{uu} +(2 F_4-3 F_5^2) A_{u})
u_{x}+(B_{u_x}-F_5 A) F_4$$$$-10 A_{u_x}
F_0+(F_{0u_x}+F_0F_3+G_6)A+B_{u_x} (G_7-F_5^2)+F_{5u}
B,\eqno(A.21)$$
$$h_7=A_{u_xu_xu_xu_x}+A (G_{1u_x}-3 F_1^2)+(-7 A_{u_xu_x}-A G_3-4 F_3 A_{u_x}) F_1$$ $$+(A_{u_xu_xu_x}+A G_1) F_3
+ A_{u_x}(F_{1u_x}+5G_1),\qquad\qquad \eqno(A.22) $$
$$h_8=[(-3 F_3 A_{u}-10 A_{uu_x}) F_1+4 A_{uu_xu_xu_x}+A G_{1u}+3 F_3 A_{uu_xu_x}+F_{1u_x} A_{u}
+4 A_{u} G_1] u_{x}$$
$$ +(-4 A_{u}-5 F_2 A-B_{u_xu_x}-F_3 B_{u_x}-4 F_5 A_{u_x}) F_1+(-4 F_3 A_{u_x}-8 A_{u_xu_x}-A G_3) F_2 $$ $$+(3
A_{uu_x}+B_{u_xu_xu_x}+A G_2) F_3+(A_{u_xu_xu_x}+A G_1)
F_5+B_{u_xu_xu_xu_x}+5 A_{u_x} G_2\qquad$$ $$-A G_5 F_1+6
A_{uu_xu_x}+A G_{2u_x} +F_{1u_x} B_{u_x}+B_{u_x} G_1+F_{2u_x}
A_{u_x}+F_{1u} A,\eqno(A.23)
$$
$$h_9=(-3 F_1 A_{uu}+3 F_3 A_{uuu_x}+6 A_{uuu_xu_x})
u_{x}^2+[(2 B_{uu_x}-3 F_5 A_{u}) F_1-3(F_3 A_{u}+4A_{uu_x}) F_2
$$ $$+(3 A_{uu}+3 B_{uu_xu_x}) F_3+F_{1u_x} B_u+A G_{2u}+12
A_{uuu_x}+F_{2u_x} A_{u}+3 F_5 A_{uu_xu_x}$$ $$+4 A_{u} G_2+4
B_{uu_xu_xu_x}] u_{x} -A G_5 F_2+(2 B_u-A G_7-F_5 B_{u_x}-4 F_4 A)
F_1-2 F_2^2 A$$
$$-(F_3 B_{u_x}+2 B_{u_xu_x}+4 F_5 A_{u_x}+5 A_{u})
F_2+(A G_4+3 B_{uu_x}-4 F_4 A_{u_x}) F_3 $$ $$-(9 A_{u_xu_x}+A
G_3) F_4 +(3 A_{uu_x}+B_{u_xu_xu_x}+A G_2) F_5+5 G_4A_{u_x}
+AF_{2u}$$
$$+F_{4u_x} A_{u_x}+B_{u_x} G_2+3 A_{uu}+F_{1u} B+6
B_{uu_xu_x}+F_{2u_x} B_{u_x}+A G_{4u_x}, \eqno(A.24)
$$
$$h_{10}=(5 A_{u_x}+F_3 A) G_6+[4 A_{uuuu_x}+F_3 A_{uuu}) u_{x}^3+(3 F_1 B_{uu}+3 F_5 A_{uuu_x}
+6 A_{uuu}+3 F_3 B_{uuu_x}$$
$$  -4 F_2 A_{uu}+6 B_{uuu_xu_x}] u_{x}^2+[3 F_5 A_{uu}-(14 A_{uu_x}+3 F_3 A_{u}) F_4-3 F_5 A_{u} F_2\qquad\quad$$
$$ +AG_{4u}+3 F_5 B_{uu_xu_x}+3 F_3 B_{uu}+4 A_{u} G_4 +12
B_{uuu_x}+F_{2u_x} B_u+F_{4u_x} A_{u}] u_{x}\qquad
$$ $$ -(3 F_2 A+6 A_{u}+A
G_5+F_3 B_{u_x}+3 B_{u_xu_x}+4 F_5 A_{u_x}) F_4-(10 A_{u_xu_x}+3
F_1 A$$
$$ +A G_3+4 F_3 A_{u_x}) F_0+3 F_5 B_{uu_x}+F_{4u} A+F_2 B_u+F_{4u_x} B_{u_x}+B_{u_x}
G_4\qquad\qquad
$$ $$-A G_7 F_2+3 B_{uu}+F_5 A G_4+A G_{6u_x}+F_{0u_x} A_{u_x}+F_{2u}
B-F_5 B_{u_x} F_2 ,\eqno(A.25)$$
$$h_{11}= (F_5 A+4 A_{u} u_{x}+B_{u_x}) G_6+(5 A_{u_x}+F_3 A) G_8+A_{uuuu} u_{x}^4
+(F_5 A_{uuu}+4 B_{uuuu_x}+F_3 B_{uuu}) u_{x}^3$$ $$ \;\;+(6
B_{uuu}-5 F_4 A_{uu}+2 F_2 B_{uu}+3 F_5 B_{uuu_x})
u_{x}^2+[F_{4u_x} B_u+A G_{6u}+3 F_5 B_{uu}+F_{0u_x} A_{u}$$
$$\;\;-(2 B_{uu_x}+3 F_5 A_{u}) F_4-(3 F_3 A_{u}+16 A_{uu_x}) F_0
]u_{x}-F_4^2 A-(F_5 B_{u_x}+A G_7) F_4 -(4 B_{u_xu_x}$$
$$+A (G_5+2 F_2)+F_3 B_{u_x}+4 F_5 A_{u_x}+7 A_{u}) F_0+(F_{0u}+G_{8u_x})A+F_{0u_x}
B_{u_x}+F_{4u} B,\eqno(A.26)
$$
$$h_{12}=(F_5 A+4 A_{u} u_{x}+B_{u_x}) G_8+B_{uuuu} u_{x}^4+F_5 u_{x}^3 B_{uuu}
-(6 A_{uu} F_0-F_4 B_{uu}) u_{x}^2$$$$+(A G_{8u}-4 B_{uu_x} F_0 -3
F_5 A_{u} F_0+F_{0u_x} B_u) u_{x}$$
$$-F_4 A F_0-F_5 B_{u_x}F_0+(-A G_7-B_u) F_0+F_{0u} B.\eqno(A.27)
$$

\end{document}